# A Multichannel RF Transceiver Array with mixed L-C loop and microstrip elements for Foot/Ankle MR Imaging at 7T

Aditya Bhosale[1], Leslie Ying[1,2], Xiaoliang Zhang[1,2]

[1]Department of Biomedical Engineering, State University of New York at Buffalo, Buffalo, NY, USA

[2]Department of Electrical Engineering, State University of New York at Buffalo, Buffalo, NY, USA

*Abstract—* **It is technically challenging to design efficient transceiver coil arrays for foot and ankle imaging at ultrahigh fields due to the irregular geometry of the anatomy. Shortened wavelength because of the high operation frequency at ultrahigh fields increases phase variation, leading to inhomogeneous B1 distribution in the imaging sample. In this work, we propose a hybrid design with mixed L-C loops and microstrip resonators for multichannel foot/ankle transceiver arrays at the ultrahigh field of 7T. The proposed transceiver array consists of 14 microstrip resonators and 5 L-C loop coils to cover the entire region of interest of the irregular-shaped foot/ankle with a relatively uniform B1 distribution at 7T. The feasibility and field behavior of the proposed design are systematically investigated numerically.**

## I. Introduction

MRI [1, 2] is extensively utilized to identify various anomalies, injuries, and illnesses in bones and soft tissues, including cartilage deterioration, bone marrow edemas, osteoarthritis, osteoporosis, and ligament/tendon injuries [3-11]. Magnetic resonance imaging has the upper hand in diagnosing disorders involving bones and soft tissues early [12-17]. Ultra-high field MRI imaging at static magnetic fields greater than or equivalent to 7T (300 MHz) improves signal-to-noise ratio (SNR), spatial, and contrast resolution [18-32] but comes with certain negatives such as difficulties in designing efficient RF coils for signal excitation and reception, inhomogeneous B1 fields, higher specific absorption rate, and image degradation abnormalities [33-45].

Most MRI scanners used for MR foot and ankle imaging are 1.5 or 3 Tesla. Still, for ultra-high-field applications such as 7T, a minimal number of array systems are available, which demands new array systems be developed. It is technically challenging to design highly efficient transceiver RF coils for foot/ankle imaging at an ultrahigh field of 7T due to the irregular shape of the anatomy. In contrast, conventional cylindrical RF coils show the limitation in filling factor and distance between coils and sample. Additionally, the shortened wavelength of the high operation frequency at ultrahigh fields poses difficulties in generating uniform B1 distribution in the conductive and high dielectric biological samples. This study aims to develop a multichannel transceiver array to cover the entire foot and ankle region with relatively uniform B1 distribution, ultimately enabling parallel imaging, parallel excitation, and B1 shimming in foot/ankle imaging at 7T. For this goal, we propose a hybrid design with mixed transceiver L-C loops and microstrip resonant elements. The hybrid multichannel transceiver array is designed in a subject-conformed shape for higher filling factor and transmit/receive efficacy. B1+ and B1- distributions and electric fields of the proposed multichannel hybrid coil array loaded with a foot/ankle phantom are investigated numerically. Potentially the proposed technique can be used to design wearable or flexible coils for foot/ankle imaging at ultrahigh fields.

Due to their distinct benefits over conventional coils, microstrips are promising in high-frequency RF coil designs for MR applications at ultrahigh fields, such as decreased radiation loss, distributed-element circuit, high-frequency capabilities, and decreased perturbation of sample loading to the RF coil in the high-frequency range [46-62]. Microstrips are purely distributed conductors made up of a narrow strip of silver or copper. A low-loss dielectric material of a specific thickness separates the ground plane from the strip conductor in the microstrip. Microstrips offer a better Q-factor and superior decoupling performance, do not require shielding, are less expensive, and are simple to manufacture. The following equation may be used to compute the frequency of a standard MTL resonator:

$$f_r = \frac{nc}{2l\sqrt{\varepsilon_{eff}}}, (n = 1,2,3...).$$

MTL resonators with capacitively terminated capacitors are identical to regular MTL resonators. The termination capacitors can be attached to one or both of the conductor's ends. The termination capacitors increase the electrical length of the microstrip line and cause it to resonate at the desired frequency.

The effective dielectric constant of the MTL is calculated using the following equation:

$$\varepsilon_{eff} = \left[1 + \frac{H_1 - H}{H}(\sqrt{\varepsilon_r} - 1)(\xi^+ - \xi^- \ln\frac{W}{H})\right]^2$$

Where

$$\xi^+ = \left(0.5173 - 0.1515 \ln\frac{H1 - H}{H}\right)^2$$

$$\xi^- = \left(0.3092 - 0.1047 \ln\frac{H1 - H}{H}\right)^2$$

The characteristic impedance of the MTL is calculated using the following equation:

$$z_0 = \frac{60}{\sqrt{\varepsilon eff}} \ln\left[\frac{H}{W}\right] \Xi + \sqrt{1 + (\frac{2H}{W})^2}$$

Where

$$\Xi = 6 + \frac{2(\pi - 3)}{exp\left(\frac{30.666H}{W}\right)^{0.7528}}$$

Where W is the width of the strip conductor, H is the height of the substrate

Finally, the resonant frequency of the MTL resonator terminated by one capacitor is calculated by using the following equation:

$$f_r = \frac{-1}{2\pi Z_0 C_t} \tan\left(\frac{2\pi l \sqrt{\varepsilon_{eff}}}{c} f_r\right)$$

And the resonant frequency of the MTL resonator terminated on both sides is calculated using the following equation:

$$f_r = \frac{(2\pi f_r Z_0)^2 C_t C_{t1} - 1}{2\pi Z_0 (C_t + C_{t1})} \tan\left(\frac{2\pi l \sqrt{\varepsilon_{eff}}}{c} f_r\right)$$

$Z_0$ is the characteristic impedance, $\varepsilon_{eff}$ is the effective permittivity, $l$ is the length of the strip conductor, and $C_t$ & $C_{t1}$ are the termination capacitors [35].

II. METHODS

The proposed array system consists of 14 capacitively terminated microstrip transmission lines[33, 35, 51, 62] and five surface loop coils closely placed on the human foot/ankle-shaped phantom. We built the foot/ankle-shaped phantom to resemble the human foot/ankle using COMSOL Multiphysics. The dimensions of the capacitively terminated MTL resonators were as follows: length of the conductor: 18 cm, width of the strip conductor: 3 mm, the width of the substrate: 1 cm, and height of the substrate: 3 mm. The hybrid array system has eight MTL resonators arranged in a circular placement to cover the ankle region of the phantom, and six MTL resonators arranged in a one-planar fashion to cover the metatarsal and phalanges region of the phantom, and five surface coils covering the heel and sides of the foot phantom. The layout and design of the simulation model are shown in the following figures. This design aims to develop an RF array system to efficiently produce field maps covering the area of interest [63-65].

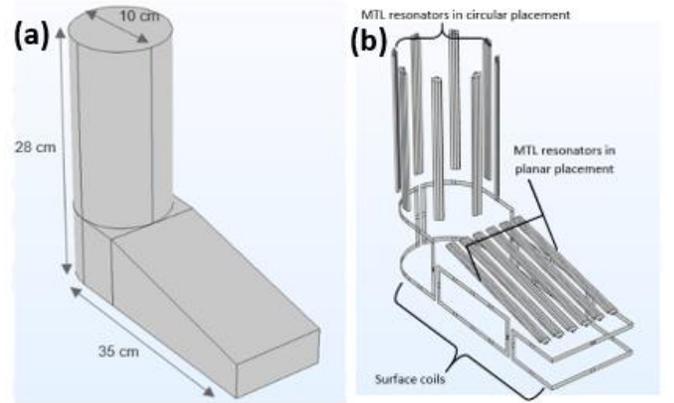

Fig. 1 (a) Geometry of the Human foot/ankle-shaped phantom and dimensions. (b) The structure of the 19-channel hybrid array system.

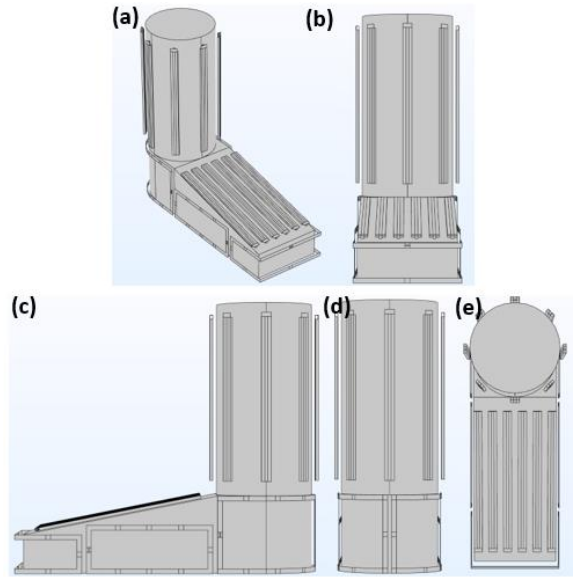

Fig. 2 The geometry of the hybrid array system loaded with the human foot/ankle-shaped phantom (a) Three-dimensional view (b) Front view (c) Side view (d) Back view (e)Top view.

## A. Array Elements

The 19-channel hybrid array comprises 14 MTL resonators (fig. 3a,3b,3c). The substrate's width, length, and thickness used in MTL resonators are 10mm, 170mm & 3mm, respectively. The width of the stripline is 3mm for all the MTL resonators. We tuned All the MTL resonators at 300MHz using two tuning capacitors ($C_t$) of 6.4pF for each element. The impedance of each MTL element was matched at 50 ohms using a matching capacitor ($C_m$) of approximately 2pF. The remaining elements in the 19-channel hybrid array comprise five surface coils. For simplicity, we have divided these coils based on their position around the human foot/ankle phantom. The five surface coils have two coils wrapping the heel region, two coils surrounding the side region of the lower foot, and one single coil wrapped around the toe region of the foot.

The width and length of the heel coils (fig. 3d,3e) are 80 mm & 100 mm, respectively. We tuned the heel coils at 300MHz using seven tuning capacitors ($C_t$) of 6.2pF and matched the impedance at 50 ohms by using a matching capacitor ($C_m$) of 15pF.

The side coils (fig. 3f,3g) surrounding the side of the lower foot had a width of 50mm & a length of 130mm. We tuned these coils using seven tuning capacitors ($C_t$) of 9pF and matched the impedance at 50 ohms by a matching capacitor ($C_m$) of approximately 18pF.

The front coil (fig. 3h) wrapped around the toe region of the foot had a bent structure and a width of 40mm & a length of 246mm. The front coil had seven tuning capacitors of 6[pF] and one matching capacitor of 19[pF] to match the coil at 300MHz/50ohms.

## B. Substrate

The material used as the substrate for the MTL resonator is commercially available as Teflon. The dielectric material properties of the substrate were $\varepsilon_r$ = 2.1, $\sigma$ = 0 s/m.

## C. Phantom

We used a human foot/ankle-shaped phantom in our simulation. The dielectric properties were set to $\varepsilon_r$ = 39, $\sigma$ = 0.49 s/m. Figure 1 shows the human foot/ankle-shaped phantom.

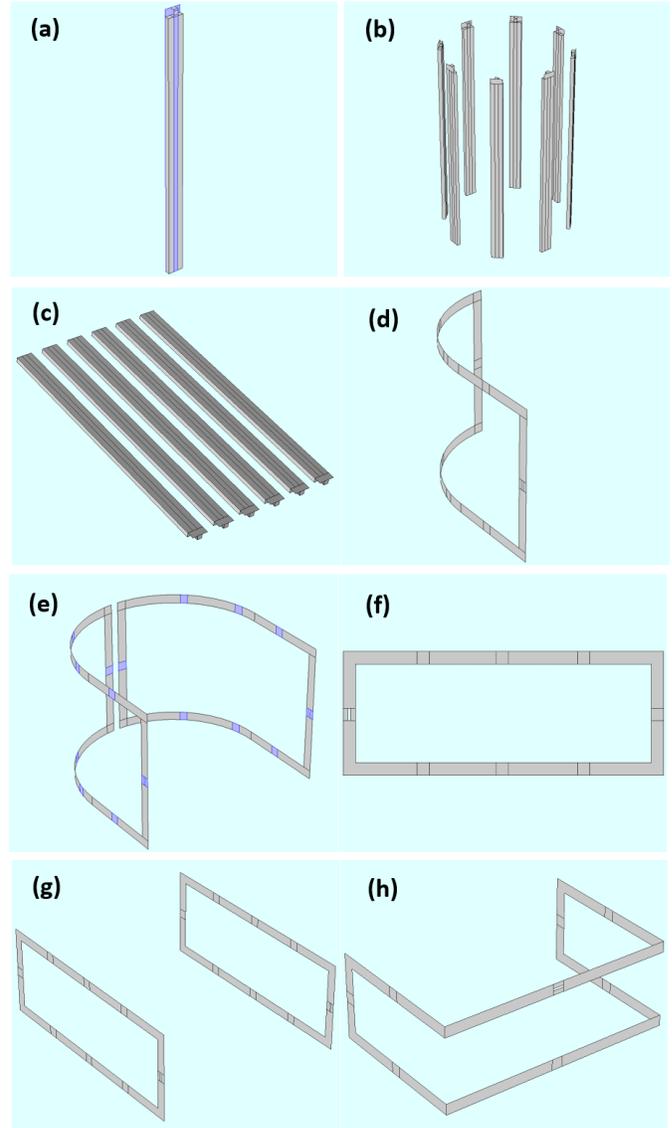

Fig. 3 The geometry of the hybrid array elements (a) MTL resonator (b) MTL resonators in circular arrangement (c) MTL resonators in planar arrangement (d) Surface coil used to wrap the heel region of the human foot phantom (e) arrangement of the heel coils in the hybrid array (f) Surface coil surrounding the side area of the lower foot (g) arrangement of the side coils in the hybrid array (h) surface coil wrapped around the toe region of the human foot.

## D. Simulations

After setting up the model and assigning the materials to respective domains, we added the Electromagnetic waves, and frequency domain physics. We used the lumped ports to feed individual channels and lumped elements to add the capacitors. The resonant frequency of the separate channel was evaluated by selecting the appropriate capacitors by computing the adaptive frequency sweep. The calculated value of the capacitors for the MTL resonators is 6 pF, while the measured value in the simulation was approximately 5.7 pF at the characteristic impedance of 50 ohms. The MTL resonators in the circular placement were excited using the same amplitude and a phase difference of $45^0$, and other channels were excited without any phase difference. We applied an extremely fine mesh setting to all domains for high-resolution field distributions. Further, we computed the frequency domain study to evaluate the electromagnetic field distribution of our proposed array system at 300MHz.

### III. Results

*A. Magnetic field distribution*

The magnetic field distribution in dB was shown using a logarithmic scale, and the following expression was used: 20*log10 (emw. normB). The magnetic field distribution is seen in the region of interest, a human foot/ankle phantom. Our findings reveal a uniform magnetic field distribution that covers the entire phantom.

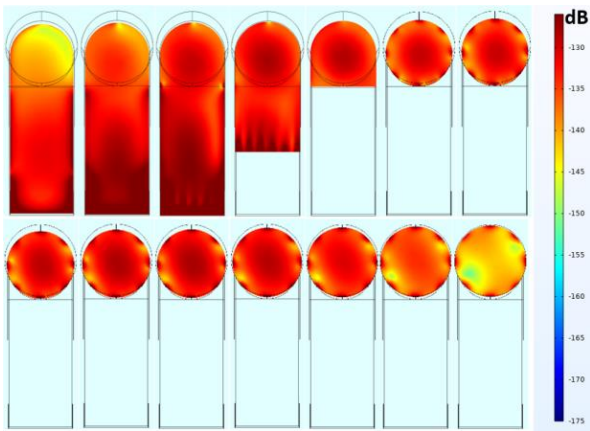

Fig. 4 Magnetic field distribution produced by the 19-channel hybrid array system in the human foot/ankle-shaped phantom at 300MHz. The figure shows the individual axial slices from the bottom to the top of the phantom.

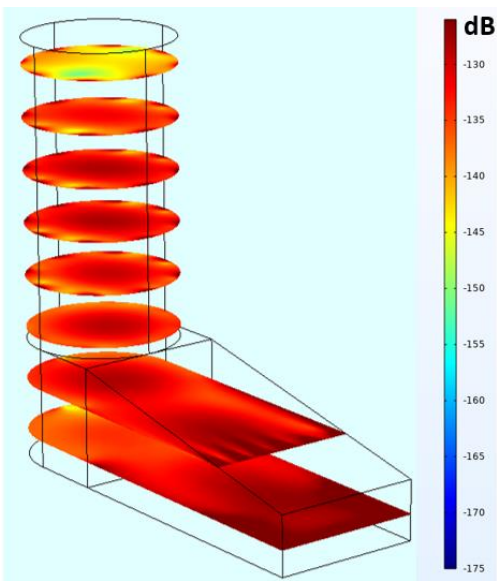

Fig. 5 Axial Multislice of the Magnetic field distribution produced by the 19-channel hybrid array system at 300MHz

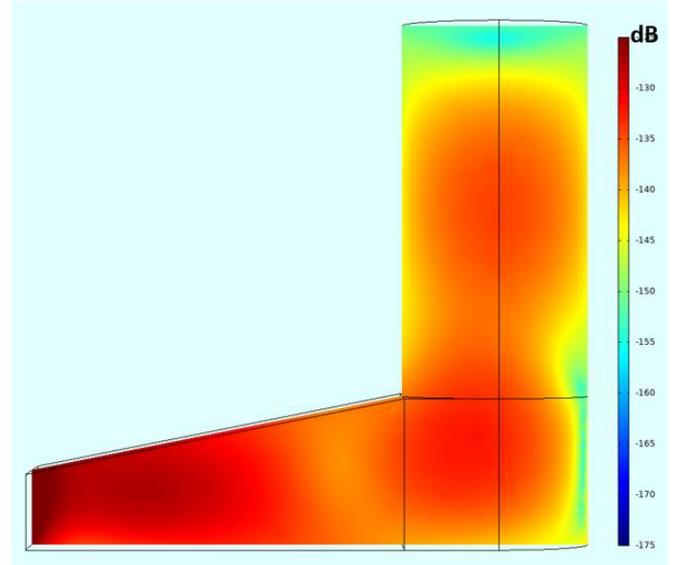

Fig. 6 Magnetic field distribution produced by the 19-channel hybrid array system in the human foot/ankle-shaped phantom at 300MHz. The figure shows a sagittal slice of magnetic fields in the phantom.

*B. Electric field distribution*

We used the linear scale to display the electric field distribution in V/m. The expression used was as follows: (emw.normE). The electric field distribution is shown in the area of interest, a human foot/ankle-shaped phantom.

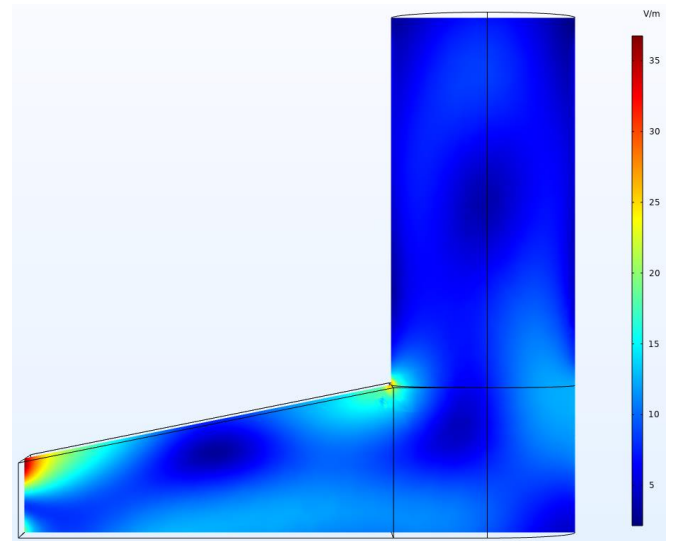

Fig. 7 Electric field distribution produced by the 19-channel hybrid array system in the human foot/ankle-shaped phantom at 300MHz. The figure shows a sagittal slice of electric fields in the phantom.

*C. B1+ & B1- field distribution*

We extracted the B1+ and B1- field maps using COMSOL Multiphysics. The field maps produced by each channel were obtained. The B1+ and B1- field evaluation is essential to determine the parallel imaging performance of our proposed array system. The B1+ and B1- field distributions are in agreement with the theory.

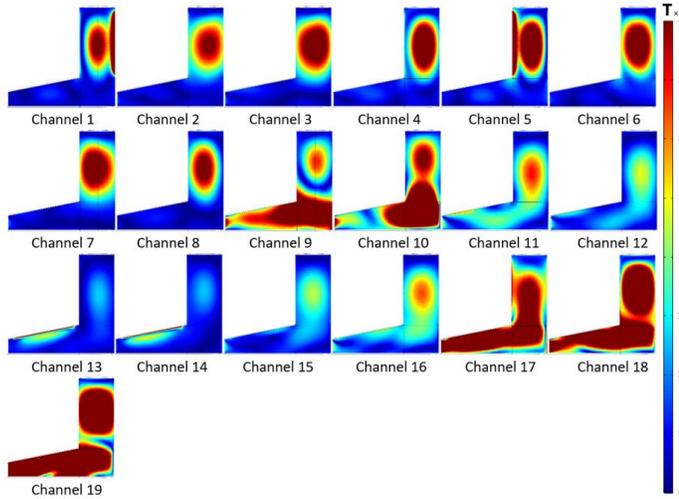

Fig. 8 The B1+ field maps produced by the 19-channel hybrid array system. The figure shows each channel's sagittal slices of the B1+ field distribution.

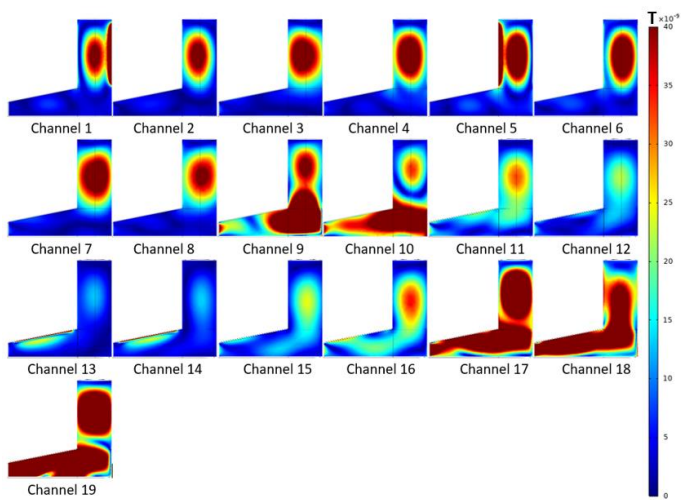

Fig. 9 The B1+ field maps produced by the 19-channel hybrid array system. The figure shows each channel's sagittal slices of the B1+ field distribution.

## IV. CONCLUSION AND DISCUSSION

In this study, we designed and simulated a 19-channel hybrid transceiver array for foot/ankle MRI consisting of L-C loops and MTL resonators. The designed hybrid transceiver array produced uniform magnetic field distribution covering the whole region of interest for the irregular-shaped anatomy of the foot and ankle. The multichannel transceiver design provides the capability of performing parallel transmission and B1 shimming, often needed in ultrahigh field MR imaging in humans. In the present design, the issue of insufficient electromagnetic decoupling among the resonant elements may remain. Possible solutions to address the decoupling include a magnetic wall decoupling technique and a high impedance resonator strategy. Although the electric field distribution produced by the multichannel hybrid array is within safe limits, the hot spots generated near the toe region of the foot phantom can possibly impose a safety concern and may need further investigation. The potential solution to reduce these hot spots includes using high dielectric materials. Future studies of a prototype based on this design through bench tests and imaging experiments can be carried out to assess further the success of this hybrid foot/ankle transceiver array system in a realistic setting.

## ACKNOWLEDGEMENT

This work is supported in part by the grant from NIH (U01 EB023829) and SUNY Empire Innovation Professorship Award.